\documentclass[a4paper]{article}  
\usepackage{graphicx}

\begin{document}
	
\begin{center}
    {\Large Geant4 FTF Model Description of the NA61/SHINE Collaboration Data on 
    	Strange Particle's Production in pp Interactions}
\end{center}

\large
\begin{center}
A. Galoyan\footnote{Veksler and Baldin Laboratory of High Energy Physics, Joint Institute for Nuclear Research,
Dubna, Moscow region, 141980 Russia},
V. Uzhinsky\footnote{Laboratory of Information Technologies, Joint Institute for Nuclear Research,    
Dubna, Moscow region, 141980 Russia}\\
\vspace{0.5cm}
on behalf of the Geant4 hadronic physics working group
\end{center}


\begin{center}
    \begin{minipage}{14cm}
		\centerline{Abstract}
 The latest data by the NA61/SHINE collaboration on inclusive distributions of $K^{*0}$ and $\phi$ meson’s, $\Xi^-$ and 
 $\bar \Xi^+$ hyperon’s production in ${\rm pp}$ interactions at $P_{lab}$ from 40 up to 158 GeV/c are considered. As it was shown before in experimental papers, Monte Carlo models -- EPOS 1.99, UrQMD 3.4 and Pythia 6, cannot describe reasonably well the data. In the presented paper, the yields of the particles are analyzed within  the Geant4 FTF model. The meson yields are proportional to the probability of strange 	quark-antiquark pair production from the vacuum, and the probability to form a vector 	meson at a given quark-anti-quark content -- $P_{Vec}$. Detailed calculations presented in our paper show that a good description of the data can be reached for $P_{Vec}$ = 0.6 for a meson with one strange quark or antiquark. For a meson contained strange quark and strange antiquark, $P_{Vec}$ = 0.7. For a combination of non-strange quark and antiquark, $P_{Vec}$ = 0.5. The description of the $\Xi^-$ and $\bar \Xi^+$ hyperons requires a change of the diquark fragmentation function into hyperons and a special treatment (rearrangement) of fragmentation of anti-diquark – diquark strings with low masses   

    \end{minipage}
\end{center}

Recently, the NA61/SHINE collaboration has published \cite{ref1,ref2,ref3} experimental data on vector mesons -- $\phi$(1020) 
and $K^{*0}$(892) + $\bar K^{*0}$(892) (referenced below as $K^{*0}$), production in ${\rm pp}$ interactions at beam momenta 
($P_{lab}$) in the laboratory system 40, 80 and 158 GeV/c. Also, the data on $\Xi^-$ and $\bar \Xi^+$ hyperons at $P_{lab}=$ 
158 GeV/c have been published \cite{ref4}. The collaboration has compared their data with EPOS 1.99 model \cite{ref5,ref6} 
predictions for inclusive particle distributions. The data for the mesons and some model calculations performed by the 
collaboration are presented in Fig.~1. As seen, the EPOS model essentially overestimates the data in the central rapidity 
region. For comparison of model predictions (EPOS 1.99 \cite{ref5,ref6}, UrQMD 3.4 \cite{ref7,ref8} and Pythia 6 \cite{ref9}) 
on total yields of the particles please refer the experimental papers \cite{ref1,ref2,ref3,ref4}. We are going to analyze the 
data within the Geant4 FTF model. All these models assume creation and fragmentation of quark strings. They differ mainly in 
string creation mechanism, though the string fragmentation is approximately the same.  

Geant4 is the well-known package for simulation of particle penetration in matter which is used in many high energy 
experiments \cite{ref10,ref11,ref12}. There is FTF (Fritiof) model in Geant4 responsible for simulations of elementary interactions. 
Thus, it is very important for various application to have a good description of experimental data with the FTF model. 
This aim can be reached by tuning of model parameters. We are using the latest version of Geant4 -- 11.0, as a default model. 

From theoretical point of view, yield of $K^{*0}$ mesons is proportional to a probability of the strange $s \bar s$ quark 
pair creation from the vacuum at quark or diquark fragmentation ($P_{s \bar s}$), and a probability of vector meson 
production -- $P_{Vec}$.  The numerical values of the probabilities are unknown for the EPOS model. In the classical paper by R.D. Field and R.P. 
Feynman \cite{ref13}, it was proposed that $P_{s \bar s}$ and $P_{Vec}$ are equal 20 and 50 \%, respectively. In the famous Pythia 6.4 model \cite{ref9}, they are 13 and 60 \% for mesons with $s$-quarks. $P_{Vec}=0.5$ for non-strange mesons. Rationales for the choice were not given. We  chose an asymptotic value of $P_{s \bar s}=12$ \% describing $K^\pm$ meson production in $\rm pp$ interactions, and $P_{Vec}=0.5$ (for all mesons) in the Geant4 FTF model. FTF calculations with these values are shown in Fig. 1 by thin solid lines. As seen, we underestimate the data in the case. Choosing $P_{Vec}=0.6$ for all mesons, we have good results (see thick solid lines in Fig. 1). Our results at $P_{lab}= 158$ GeV/c are very close to the EPOS ones. This points out that $P_{Vec}$ in the EPOS model is about 3/4 (maximum allowed value). At our choice of $P_{Vec}=0.6$, we overestimate $\pi^\pm$ meson production in pp interactions. Thus, we use $P_{Vec}= 0.6$ only for mesons with one strange quark or antiquark. The analogous approach is used in the Pythia 6.4 model (see Ref. \cite{ref9}, page 490, parameters PARJ(11), PARJ(12) and PARJ(13)).

\begin{figure}[hbt]
	\centering 
	\resizebox{6in}{1.5in}{\includegraphics{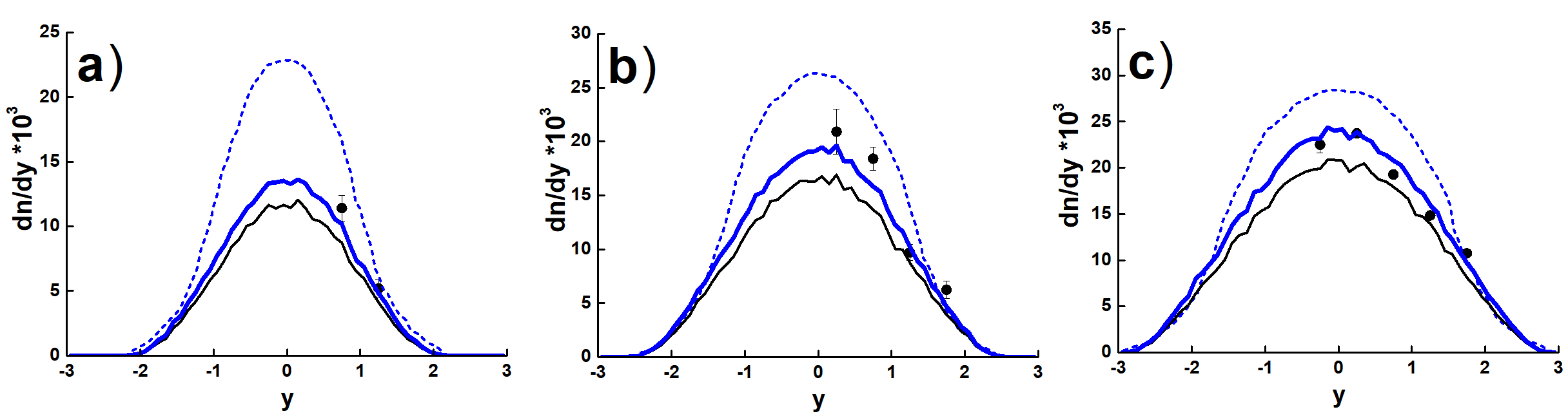}}
	\caption{Rapidity distributions, $dn/dy$,  of $K^{*0}$ mesons for three beam momenta in $\rm pp$ interastions. 
		Black points are data measured by the NA61/SHINE collaboration with statistical errors only. Dashed lines 
		are the predictions of the EPOS model obtained by the collaboration. Predictions of the Geant4 FTF 
		model with $P_{Vec} = 0.5$ (default) and $0.6$ are shown by thin and thick lines, respectively.
	}
	\label{Fig1}
\centering 
\resizebox{6in}{1.5in}{\includegraphics{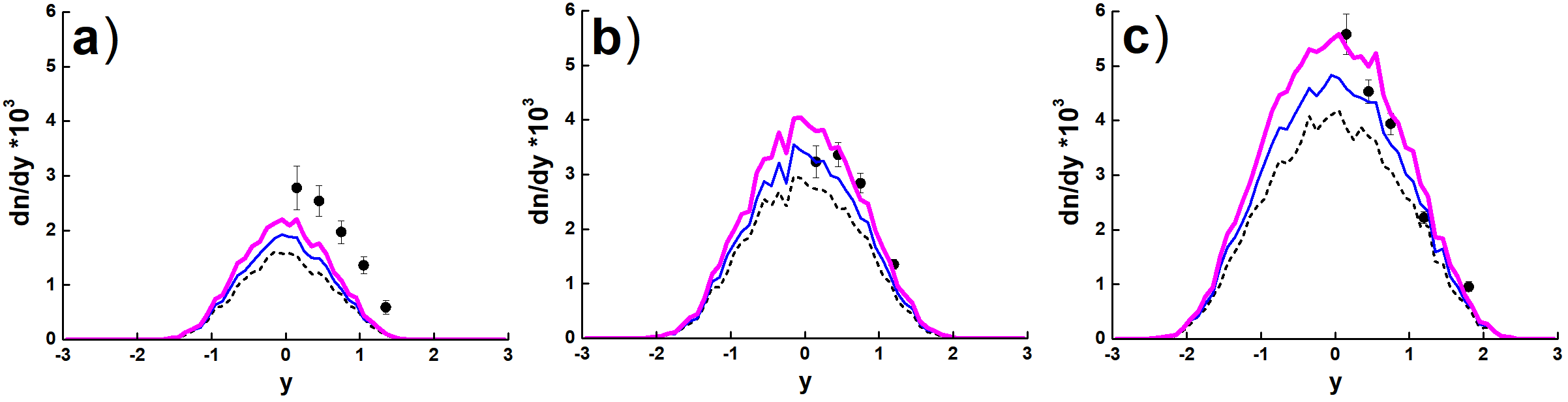}}
\caption{Rapidity distributions, $dn/dy$,  of $\phi$ mesons for three beam momenta. 
	Black points are data measured by the NA61/SHINE collaboration with statistical errors only. 
	Predictions of the Geant4 FTF model with $P_{Vec}$ = 0.6 and 0.7 are shown by thin and thick lines, 
	respectively. Dashed lines are calculations for $P_{Vec}$ = 0.5 .
}
\label{Fig2}
\end{figure}

The weak energy dependence of the EPOS results shows also that there is, probably, no energy dependence of $P_{s \bar s}$ in the model. It is obvious that $P_{s \bar s}$ must depend on a string mass. If the mass is below two $K$-meson masses, 
a production of two strange mesons is impossible, $P_{s \bar s}=0$. Thus, we introduced a dependence of $P_{s \bar s}$  on string mass: $P_{s \bar s}= P^{ass}_{s \bar s}\ [1 – (M_{th}/M_{str})^{2.5}]$, which leads to a good energy dependence of 
$K$-meson production. Here $P^{ass}_{s \bar s}=$ 12 \% is the asymptotical value of the probability, $M_{th}$ is 
the threshold mass value about 1.25 GeV, and $M_{str}$ is a string mass. 

Proposed values of the parameter ($P_{Vec}$) do not allow a satisfactory description of the $\phi$ meson production. 
However, changing $P_{Vec}$ value to 0.7 for mesons with strange quark and antiquark ($\phi$), better results can 
be obtained (see Fig. 2).

As seen, our Geant4 FTF model underestimates only the data at $P_{lab}=$ 40 GeV/c. At the moment it is not clear 
what has to be done to erase this drawback? An increasing of the vector meson production probability will lead 
to overestimation of the data at higher energies. The same results can be obtained at an increasing of $P_{s \bar s}$.
Additional to this, the yields of $K^\pm$ mesons will be increased also. Maybe, it is needed to consider an alternative mechanism of $\phi$ meson production at low energy. It is possible, of course, that the corresponding experimental data 
may be changed.

More complicated situation takes place with a description of the $\Xi^-$ and $\bar \Xi^+$ hyperon data shown in Fig.~3. 
at $P_{lab}= 158$ GeV/c. As seen, the EPOS model underestimates the hyperon production in the central region.  
The UrQMD model describes well the $\Xi^-$ yield, and essentially overestimates the $\bar \Xi^+$ yield. Predictions of 
the Geant4 FTF model are closed to UrQMD ones in this case. Detailed analysis shows that $\bar \Xi^+$ hyperons are 
mainly produced at last $ss$--$\bar s \bar s$ string's decays at the given energy. 

\begin{figure}[hbt]
	\centering 
	\resizebox{6in}{2in}{\includegraphics{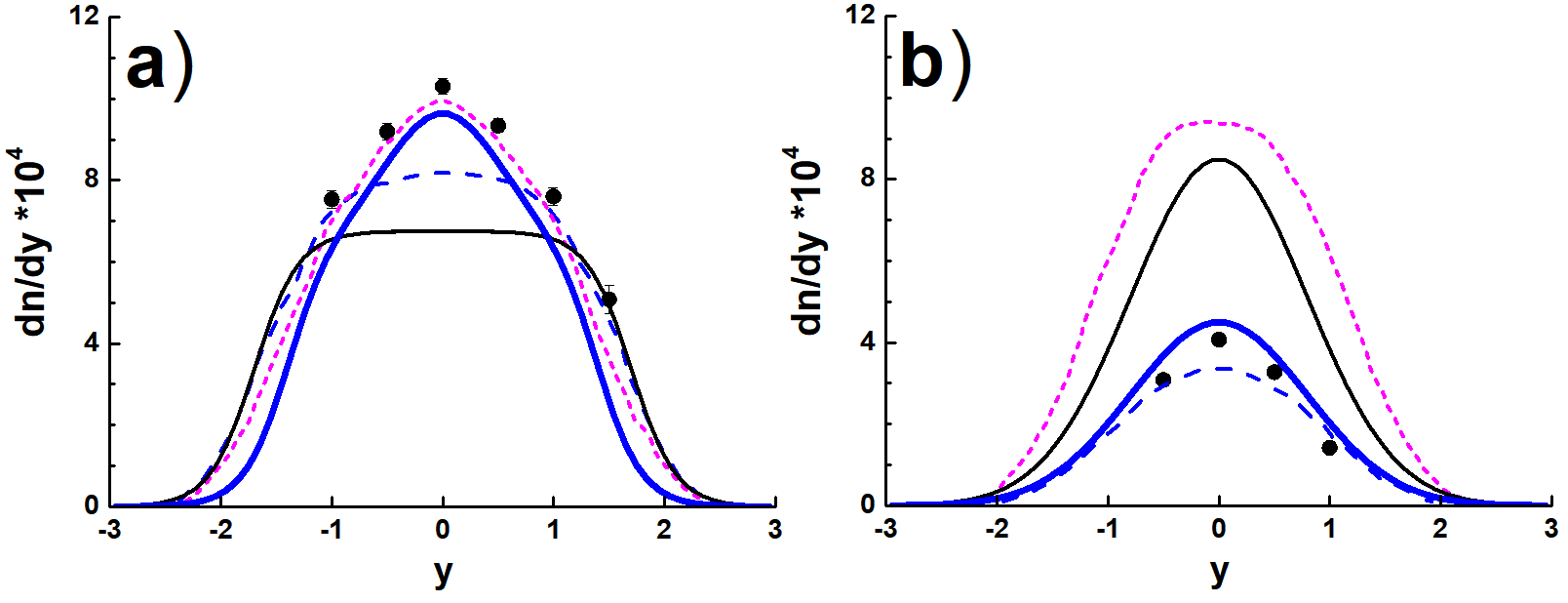}}
	\caption{Rapidity distributions, $dn/dy$,  of $\Xi^-$ and $\bar \Xi^+$ hyperons at $P_{lab}=$ 158 GeV/c (figs.
		a and b, correspondently). Black points are data measured by the NA61/SHINE collaboration with statistical 
		errors only \cite{ref4}. Predictions of the standard Geant4 FTF model are shown by thin solid lines. 
		The thick solid line (blue) for $\Xi^-$ hyperons shows the results with the fragmentation function $f(z) \propto (1-z)^\alpha$. The analogous line for $\bar \Xi^+$ hyperons was obtained assuming rearrangement of strange 
		quarks and strange anti-quarks at last decay of the diquark – anti-diquark strings ($s s$ -- $\bar s \bar s$). The short-dashed lines are UrQMD model predictions, and the long-dashed lines show EPOS results \cite{ref4}.  
	}
	\label{Fig3}
\end{figure}

It is assumed in the Fritiof (FTF) model \cite{ref14,ref15} that two quark-diquark strings are produced in an inelastic 
pp interaction at sufficiently high energies. The strings fragment emitting hadron. A string with low mass at the end 
of the fragmentation process decays into two last hadrons. These take place in the UrQMD and Geant4 FTF models. 
More complicated hypothesis is used in the EPOS model.

In the last quark-antiquark string decay two meson are produced -- $q+\bar q \rightarrow M_1(q \bar q_s)\ +\ M_2(q_s \bar q)$, 
where $q_s$ and $\bar q_s$ are a pair of see quark and antiquark created from the vacuum in the field of the string. 
A meson ($M$) can be pseudoscalar or vector meson. In the last quark-diquark string decay, a baryon ($B$) and a meson 
are produced -- $q_1+q_2\ q_3 \rightarrow M(q_1 \bar q_s)+B(q_s q_2 q_3)$. There can be 2 possibilities in a last 
anti-diquark -- diquark string decay: a) $\bar q_1\ \bar q_2\ +\ q_3\ q_4 \rightarrow \bar B(\bar q_1\ \bar q_2\ \bar q_s)\ 
+\ B(q_s q_3\ q_4)$; b) (rearrangement) $\bar q_1\ \bar q_2\ +\ q_3\ q_4 \rightarrow M_1(\bar q_1\ q_3)\ +\ M_2(\bar q_2\
 q_4)$. The process "a" allows two baryon production in an anti-p p annihilation, for example. It "works" quite well for 
 all other hadronic reactions. But for a description of the $\bar \Xi^+$ hyperon's production, we have to assume 
 the process "b" only for a last decay of a string with strange diquark and strange anti-diquark: $\bar s_1\ \bar s_2\ 
 +\ s_3\ s_4 \rightarrow M_1(\bar s_1\ s_3)\ +\ M_2(\bar s_2\ s_4)$. The results of the hypothesis can be seen in Fig. 3b.

If the hypothesis is true, we can expect that in $\rm pp$ interactions two neighboring $\phi$ mesons in an event can be
produced. An alternative hypothesis of the two $\phi$ meson production can be a correlated creation of two pairs of 
$s \bar s$ from the vacuum. We hope, the future experimental studies will shed light on the subject.

The form of the $\Xi^-$ rapidity spectrum depends of the fragmentation functions. It is too complicated to change the functions using standard possibility of the LUND string fragmentation algorithm \cite{ref16} implemented in UrQMD and Geant4 FTF models. Thus, we use for di-quark fragmentation into baryon the recipe of the Quark-Gluon String Model (QGSM) \cite{ref17,ref18,ref19}, namely, $f(z) \propto z^\alpha$. For a di-quark fragmentation into $\Xi^-$ hyperons, we have to use 
the function $f(z) \propto (1-z)^\alpha$ which allowed to obtain the hump in the central rapidity region. 

\centerline{\bf Conclusion}
Tuning the main parameters of the string fragmentation model -- $P_{s \bar s}$ and $P_{Vec}$, it is possible to reach 
a good description of the experimental data on the vector meson production in $\rm pp$ interactions, at least, in the Geant4 
FTF model. For description of the $\Xi^-$ and $\bar \Xi^+$ hyperon’s data it is needed to specify more exactly string
fragmentations into the hyperons. 

The authors are thankful to M.A. Ivanov (BLTP, JINR) and A. Ribon for useful considerations of the paper.

\end{document}